\documentclass{article}

\usepackage[nonatbib, final]{neurips_2021_ml4ps}
\usepackage[square,numbers]{natbib}
\usepackage[utf8]{inputenc} % allow utf-8 input
\usepackage[T1]{fontenc}    % use 8-bit T1 fonts
\usepackage{hyperref}       % hyperlinks
\usepackage{url}            % simple URL typesetting
\usepackage{booktabs}       % professional-quality tables
\usepackage{amsfonts, bm}       % blackboard math symbols
\usepackage{nicefrac}       % compact symbols for 1/2, etc.
\usepackage{microtype}      % microtypography
\usepackage{xcolor}         % colors
\usepackage{graphicx}
\usepackage{siunitx}

\usepackage{marginnote}

\title{Probabilistic neural networks for predicting energy dissipation rates in geophysical turbulent flows}

\author{% 
Sam F. Lewin \\ 
University of Cambridge \\
\texttt {\href{mailto:sl918@cam.ac.uk}{sl918@cam.ac.uk}} \\
\And
Stephen M. de Bruyn Kops \\
University of Massachusetts Amherst \\
\texttt{\href{mailto:debk@umass.edu}{debk@umass.edu}} \\
\And
Gavin D. Portwood \\
Lawrence Livermore National Laboratory \\ 
\texttt{\href{mailto:gdportwood@gmail.com}{gdportwood@gmail.com}}\\
\And
Colm-cille P. Caulfield \\
University of Cambridge \\
\texttt{\href{mailto:cpc12@cam.ac.uk}{cpc12@cam.ac.uk}} \\
}

\begin{document}

\maketitle

\begin{abstract}
Motivated by oceanographic observational datasets, we propose a probabilistic neural network (PNN) model for calculating turbulent energy dissipation rates from vertical columns of velocity and density gradients in density stratified turbulent flows. We train and test the model on high-resolution simulations of decaying turbulence designed to emulate geophysical conditions similar to those found in the ocean. The PNN model outperforms a baseline theoretical model widely used to compute dissipation rates from oceanographic observations of vertical shear, being more robust in capturing the tails of the output distributions at multiple different time points during turbulent decay. A differential sensitivity analysis indicates that this improvement may be attributed to the ability of the network to capture additional underlying physics introduced by density gradients in the flow.
\end{abstract}

%The PNN model is able to accurately reproduce important small-scale spatial structure in the flow whilst also providing useful uncertainty estimates alongside predictions.

\section{Introduction}\label{sec:intro}

Turbulence, as characterised by chaotic, irregular fluid motions on a range of spatial scales, is present in the vast majority of real-world fluid flows. An important characteristic of turbulent flows is their ability to mix fluid much more vigorously than non-turbulent ones, leading to effective transfer of properties such as heat and momentum. Quantifying these transfers often relies on examining the `cascade' of energy, which is transferred from motions at larger scales to the smaller scales, and eventually diffused by viscous friction as heat. Most ocean flows exhibit turbulence, which facilitates in particular the vertical mixing of heat, carbon, nutrients and other important properties that influence global energy budgets and biological activity \cite{Wunsch04}. Accurately capturing mixing processes in climate models requires knowledge of the small-scale flow structure produced by a turbulent cascade of eddies \cite{Gregg18}. The ocean is stably stratified according to density, which means deeper waters are denser and hence less buoyant than those above. This feature means that the small-scale turbulent structure is often highly heterogeneous, meaning accurate models of turbulent mixing are difficult to obtain.

One of the most important properties characterising turbulence is the dissipation rate $\varepsilon$, i.e. the rate at which kinetic energy of the turbulence is converted into internal heat energy by viscous forces at the smallest scales. Calculating the true value of $\varepsilon$ from measurements requires that 9 spatial velocity derivatives be resolved simultaneously, therefore most observational studies appeal to simple surrogate models that depend on 1 or 2 velocity derivatives, based on the simplifying assumption that turbulence is homogeneous and isotropic, that is, statistical properties of the flow are independent of position and direction. However, these assumptions are generally invalid in stratified turbulent flows, as can be tested using high-fidelity direct numerical simulations (DNS) which aim to resolve fully all of the scales of structure in the flow \cite{dBK19}. 

 The wealth of data available from DNS may be readily utilised by machine learning methods, which have become a popular choice for reduced order modelling of fluid flows due to their inherent ability to capture complex spatio-temporal dynamics \cite{Brunton20,Portwood21, Salehipour19} and reveal insights into flow physics \cite{Couchman21, Callaham21}. Recently, practical uncertainty estimates available from probabilistic neural networks have been gaining increasing popularity in geophysical and environmental applications \cite{Maulik20, Barnes21}.  In this work, we construct a probabilistic neural network (PNN) model for predicting dissipation rates $\varepsilon$ in stratified turbulent flows from statistics that are readily obtainable from many oceanographic datasets. Our model is unique in that it is able to utilise properties of both the density and velocity structure of the flow to make predictions that are accurate and robust even once turbulence has started to decay. We are particularly interested in model interpretability, exploring the ability of the PNN to capture physical relationships in the data in order to predict the particular state of the turbulence from a given column of data. 

\section{Problem outline}\label{sec:prob_outline}
The local turbulent dissipation rate $\varepsilon$ is formally defined as 
\begin{equation}
    \varepsilon = 2\nu s_{ij}s_{ij};\ \ s_{ij}=\frac12 \left(\frac{\partial u_i}{\partial x_j} +\frac{\partial u_j}{\partial x_i}\right),
\end{equation}
where $\nu$ is the kinematic viscosity and $(u_1,u_2,u_3)=(u,v,w)$ are fluctuating turbulent velocities, which are defined by separating the corresponding velocity fields into a mean (spatially or temporally averaged) part and its fluctuation (for details, see \cite{Pope00}). Assuming that the turbulence is both homogeneous and isotropic, it can be shown that $\varepsilon$ may be calculated from vertical shear $S^2 = (\partial u/\partial z)^2 + (\partial v/\partial z)^2$ alone as 
\begin{equation} \label{eq:surrogate} \varepsilon = \frac{15\nu}{4} S^2. \end{equation}
This baseline surrogate model is commonly used to calculate $\varepsilon$ from vertical profiles of shear in the ocean (see, e.g. \cite{Wesson94}), and we will use it a benchmark model for comparison. It is now appreciated that, for stratified turbulence, local vertical density gradients $\partial \rho/\partial z$, which are also readily available in many oceanographic datasets, may under some circumstances be correlated with $\varepsilon$ \cite{Caulfield21, Portwood16}. We attempt to exploit this feature by building a machine learning model for calculating local values of $\varepsilon$ that, in the style of oceanographic observations, takes vertical profiles of $S^2$ and $\partial \rho/\partial z$ as inputs, accounting for interactions between the two on a range of spatial scales. 

\section{Methods}\label{sec:methods} 
\subsection{Dataset}\label{subsec:dataset}

We train and test our model using 5 snapshots in time from a high-resolution simulation of decaying strongly stratified turbulence, obtained by numerically solving the three-dimensional Navier-Stokes equations that govern the flow (see \cite{dBK19} for details). The simulation is designed to reproduce flow conditions similar to those in the ocean, in a cuboidal domain approximately \SI{5}{\metre} tall and \SI{10}{\metre} wide in the horizontal directions, similar to the recent work of \citet{Taylor19}.  Flow snapshots are taken at various times $T = 1,2,4,6,8$ during the decay period corresponding to increasing multiples of a natural timescale imposed by the stratification (formally referred to as the buoyancy time period \cite{dBK19}). These times cover an extent over which the flow evolves from an energetic, almost isotropic turbulent state at $T=1$ to a layered regime at $T=8$ where viscosity dominates and turbulence is significantly decayed. Fields have been evenly sparsed by a factor of 2 for efficient computation, so that the total snapshot size is $N_x\times N_y \times N_z $, where $N_x/2=N_y/2=N_z=2048$. 

For training, we randomly sample 12000 vertical columns with a height of $500$ grid points from each 3D snapshot, so that the largest vertical structures are comfortably contained within each column. Letting $X=S^2$ and $Y = \partial\rho /\partial z$, inputs to the model are the values of $X$ and $Y$ at each grid point within a given vertical column $i$: $\mathbf{X}^i =[X_1^i, \ldots,  X_{500}^i]$ and $\mathbf{Y}^i=[Y_1^i, \ldots,Y_{500}^i ]$. Note that the model is trained on data from all time steps simultaneously. Outputs are column-wise values of dissipation $\bm{\varepsilon}^i = [\varepsilon_1^i, \ldots, \varepsilon^i_{500}].$ For testing, we use a vertical slice from each 3D snapshot of $500\times 500$ grid points which corresponds to feeding in 500 adjacent vertical columns (all independent from the training samples). This enables better visualisation of the ability of the model to reproduce spatial structure in the flow.

\subsection{Network overview}\label{subsec:network}
\begin{figure}
  \centering
  \includegraphics[width=0.75\textwidth]{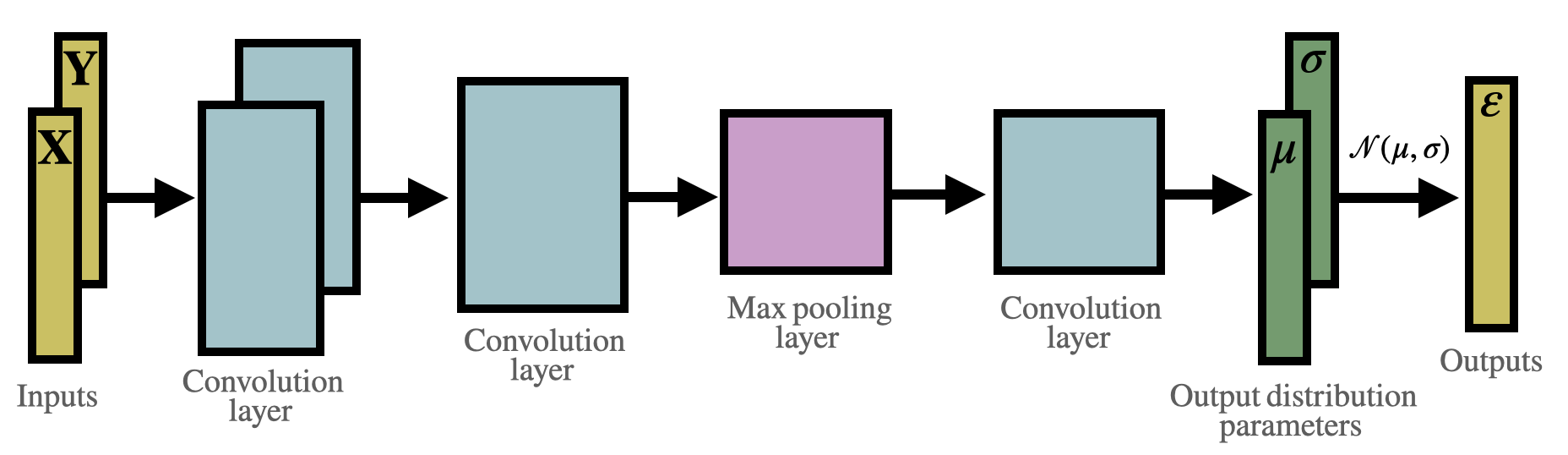}
  \caption{Schematic showing model architecture. The two input columns are first convolved separately before being combined in the second layer of the network. Model outputs are sampled from a normal distribution with learnt mean and variance that depend on the inputs.}\label{fig:model_architecture}
\end{figure}
Neural network outputs are most often deterministic, providing a single output for each input. A probabilistic nature (and hence uncertainty estimates) may be incorporated by sampling outputs from a probability distribution whose parameters are learnt by the network, resulting in a PNN. We sample each output value of $\varepsilon$ from an independent Gaussian distribution with mean $\mu$ and variance $\sigma$ so that the network can be thought of as a mapping
\begin{equation} F: (\mathbf{X}, \mathbf{Y}) \mapsto (\mu_1, \sigma_1, \cdots, \mu_{500}, \sigma_{500}). \end{equation}
Note that since the $\mu_j$ and $\sigma_j$ are independent and functions of the inputs, this does not necessarily constrain the column-wise distribution of outputs to be Gaussian. The input size is $2\times 500$ corresponding to the column heights of $\mathbf{X}$ and $\mathbf{Y}$ whilst the output size is $1 \times 1000$ corresponding to a mean and variance for each grid point in the output column. We use a convolutional structure for our network whose architecture is outlined in figure \ref{fig:model_architecture}. All convolutional layers have kernels of size 3 to appropriately capture fine-scale flow structure, leading to a total of approximately $10^7$ trainable parameters. To handle a distribution of outputs, we train the PNN by minimising a negative log-likelihood loss function 
\begin{equation}\mathcal{L} = -\sum_{i,j} \log p^i_j, \end{equation}
where $p^i_j$ is the value of the probability density function of a normal distribution with mean $\mu^i_j$ and standard deviation $\sigma^i_j$ evaluated at $\varepsilon^i_j$. Here $i$ and $j$ represent the sample number and column index respectively. We use the Adam optimizer with a learning rate of 0.005 for performing stochastic gradient descent. The model is trained for 30 epochs, taking approximately 5 minutes on a single NVIDIA Volta V100 GPU. More details can be found on the model GitHub page \url{https://github.com/samlewin/PNN_dissipation}. 
% \gdp{There needs to be a little more detail on the NN architecture and training. At minimum, state the total number of unknowns, your optimizer (adam?), number of epochs, approximate training time. Refer the reader to your github code for more details on architecture. Reduce details in first paragraph in 3.1 if you need more space.}

% \begin{ack}
% Use unnumbered first level headings for the acknowledgments. All acknowledgments
% More information about this disclosure can be found at: \url{https://neurips.cc/Conferences/2021/PaperInformation/FundingDisclosure}.

% Do {\bf not} include this section in the anonymized submission, only in the final paper. You can use the \texttt{ack} environment provided in the style file to autmoatically hide this section in the anonymized submission.
% \end{ack}

\section{Results}\label{sec:results}
\subsection{Evaluation on the test set}\label{subsec:evaluation}
\begin{figure}
  \centering
  \includegraphics[width=\textwidth]{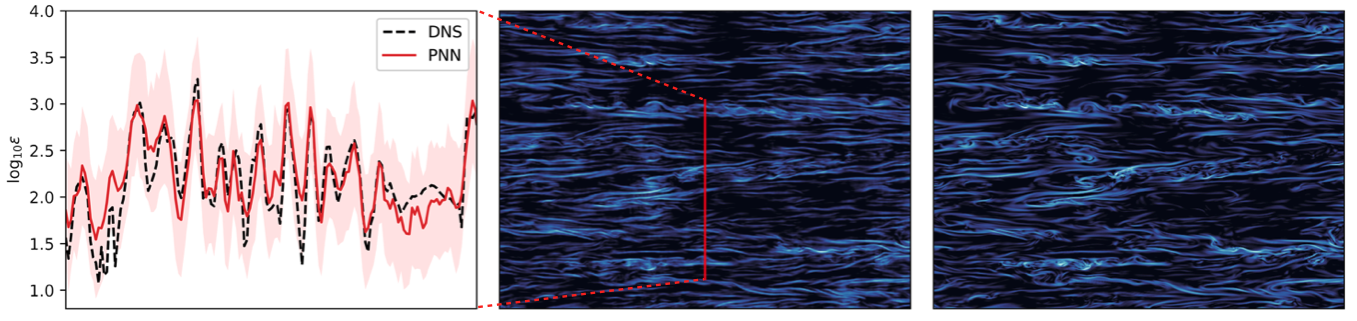}
  \caption{Left: Mean PNN predictions on an indicated section of a vertical column at time step $T=8$, with shading representing standard error uncertainties. Centre and right: Contour plots showing mean PNN predictions and true values of $\varepsilon$ respectively on a vertical slice (500 adjacent vertical columns) from time step $T=8$. Axes and colour maps use a logarithmic scale for visual clarity.}\label{fig:2}
\end{figure}

For each time step, we pass the test data through the PNN 100 times to obtain an output distribution of predicted values of $\varepsilon$. Figure \ref{fig:2} displays the distribution of output values for inputs in a section of a vertical column at time step $T=8$, demonstrating how the model is able to provide reasonable uncertainty estimates along with predictions, whilst maintaining spatial structure. Mean predictions in general capture spatial variability at the smallest vertical scales in the flow and the distinct layered structure induced by the presence of the stratification is accurately reproduced as can be seen by comparing the PNN predictions and true values displayed in the centre and right panels.

\begin{figure}
  \centering
  \includegraphics[width=\textwidth]{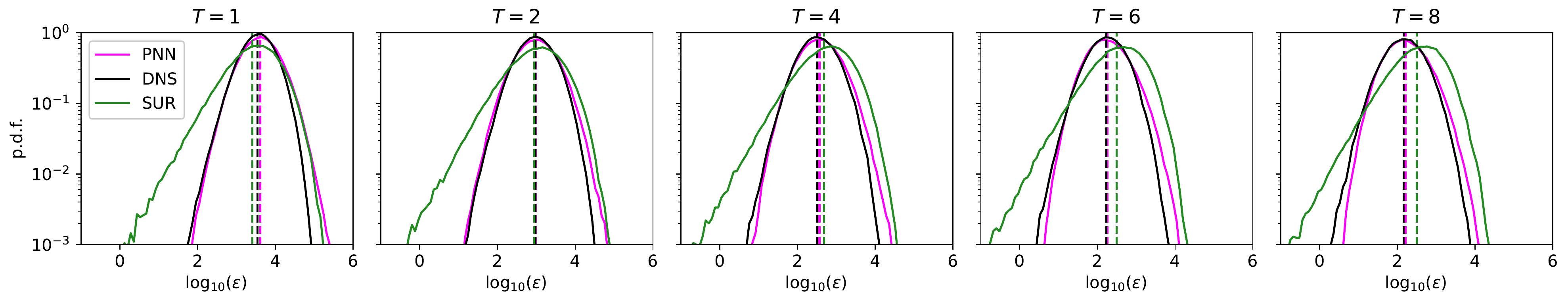}
  \caption{Normalised frequency distributions over the entire test dataset at each time step, created from a sample of PNN outputs. Also plotted are the frequency distributions of true values (DNS) and values predicted by the surrogate model (SUR) defined in (\ref{eq:surrogate}). Vertical dashed lines denote the corresponding means. Note the logarithmic axes.}\label{fig:3}
\end{figure}

Figure \ref{fig:3} shows the distribution of predicted values taken from a single sample for each time step, compared to the true distribution of $\varepsilon$ and the distribution obtained from using the baseline surrogate model defined in (\ref{eq:surrogate}). Distribution means are also indicated. It is seen that the PNN is able to capture the shape and temporal evolution of the distribution of $\varepsilon$ remarkably well, with mean values practically indistinguishable from the true value. Moreover, the PNN greatly outperforms the surrogate model both in terms of the shape of the tails and mean predicted value, particularly for later time steps as the turbulence decays and the underlying assumptions of homogeneity and isotropy break down. We found that the accuracy on the tails of distributions in particular was improved when using a probabilistic network as opposed to a regular feed-forward neural network, without a loss of accuracy in the mean value predictions. Thus there is an indication that the network uncertainty may be capturing some of the small-scale spatial intermittency of the flow. 
\subsection{Intepreting the PNN}\label{subsec:interp}
\begin{figure}
  \centering
  \includegraphics[width=\textwidth]{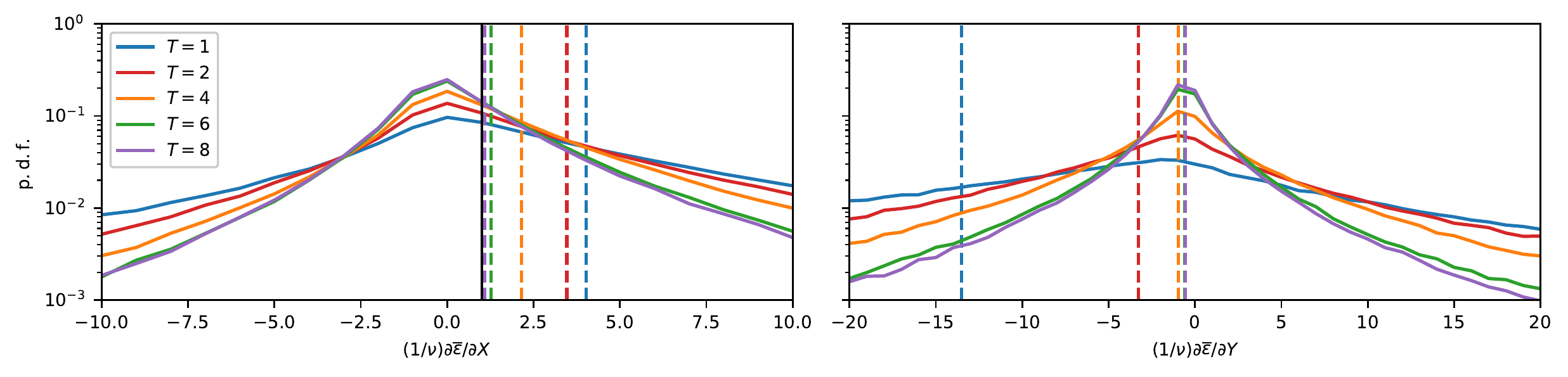}
  \caption{Distribution of derivatives of mean column outputs with respect to local inputs $X_j^i$ (left) and $Y_j^i$ (right) at each time step, normalised by $1/\nu$. Vertical dashed lines denote the corresponding distribution means, whilst the solid black line in the left-hand panel represents a reference value of 1.}\label{fig:4}
\end{figure}

It is natural to ask which aspects of the PNN enable it to make better predictions than the surrogate model (\ref{eq:surrogate}), and whether analysis of the model structure reveals the ability to capture underlying physics. To this end, we perform differential sensitivity analysis of the model by taking a derivative of the mean output values over each column $\overline{\varepsilon}$ with respect to local inputs (c.f. \citet{Portwood21}), normalising each value by $1/\nu$. This produces a set of gradients with the same shape as the inputs for each column. We then create histograms showing the distributions of these gradients over the test set at each time step for shear ($X_j^i$) and density gradient ($Y_j^i$) inputs, as shown in figure \ref{fig:4}. In general, it is seen that the distributions of gradients with respect to both shear and density gradients narrow as turbulence decays, whilst mean values decrease for shear gradients and increase for density gradients. Gradients with respect to shear inputs appear to approach a mean value of unity, which is in line with the theoretical and experimental predictions of a quasi-horizontal turbulent regime \cite{Hebert06, Riley03}. It is also seen that, at least to first order, density gradients are more negatively correlated with regions of larger dissipation for earlier time steps before the distribution shifts to the right so that, as turbulence decays, there is an increased frequency of regions where positive (unstable) density gradients are associated with larger values of dissipation, a behaviour that is commonly associated with simulations of layered strongly stratified turbulence \cite{Portwood16}. Unlike the baseline surrogate model (\ref{eq:surrogate}), the network is able to `evolve' by using the structure of the inputs to infer some of the leading order differences in flow behaviour between each point during the turbulent decay, without explicit prior knowledge of the time step.

\section{Conclusions}\label{sec:conclusions}
We have proposed a PNN model for computing local dissipation rates in stratified turbulent flows from vertical columns of shear and density gradients, as would normally be available in oceanographic observational datasets. The PNN model, which does not require any simplifying assumptions about the turbulence, robustly  outperforms a benchmark surrogate model commonly used in oceanographic practice, whilst also providing reasonable uncertainty estimates alongside predictions. Examination of the first order behaviour of the PNN revealed that the model is able to capture the evolution of the underlying physics of the turbulence without requiring prior knowledge of the particular time point during turbulent decay. This feature highlights some of the promise of machine learning methods for reduced order modelling of transient turbulent flows. Testing our model on ocean-realistic DNS of stratified turbulence is the first step towards practical application to oceanographic data. We note that observational data provides an additional challenge due to the fact that no exact measurements of dissipation exist for comparison, meaning that testing the model on data from a range of turbulent flow simulations, in addition to simulation data with artificially added noise, will be important for justifying the practical use of the PNN. We anticipate that the probabilistic nature of the model and associated uncertainties may prove especially useful during this process.

\section*{Broader impact}\label{sec:broader_impact}
Accurate models of small-scale turbulence statistics such as the dissipation rate are of crucial importance for many environmental and engineering applications. The model introduced in this work is more accurate than the current most widely used theoretical models when evaluated on data designed to represent oceanographic observational measurements, and, once trained, has no additional computational cost. Whilst this study focuses on application to data from simulations rather than observations, the results demonstrate our model is a promising proof-of-concept as a tool for the oceanographic community, where improved models of turbulent energy dissipation will increase our understanding of vertical mixing processes which have a large influence on our changing climate. The authors are not aware of any negative impacts or ethical concerns related to the work. 
\bibliography{neurips_2021}

\section*{Acknowledgements}
We would like to thank two anonymous reviewers whose constructive comments have improved the quality of this paper. An award of computer time was provided by the INCITE program. This research also used resources of the Oak Ridge Leadership Computing Facility, which is a DOE Office of Science User Facility supported under Contract DE-AC05-00OR22725. S. Lewin is supported by an Engineering and Physical Sciences Research Council DTP studentship from UKRI.

\end{document}